\documentclass[%
 aip,
 jmp,%
 amsmath,amssymb,
preprint,%
]{revtex4-1}

\usepackage{graphicx}
\usepackage{dcolumn}
\usepackage{bm}
\usepackage{epstopdf}
\usepackage{amssymb}
\usepackage{color}
\usepackage[colorlinks=true, letterpaper=true, pdfstartview=FitV, linkcolor=blue, citecolor=blue, urlcolor=blue]{hyperref}
\usepackage{lineno}

\begin{document}

\title{Stable Ferromagnetism and High Curie Temperature in VGe$_{2}$N$_{4}$}
\author{Yingmei Li}
\author{Yong Liu}\email{yongliu@ysu.edu.cn}
\affiliation{State Key Laboratory of Metastable Materials Science and Technology \& Key Laboratory for Microstructural Material Physics of Hebei Province, School of Science, Yanshan University, Qinhuangdao 066004, China }

\begin{abstract}
The discovery of monolayer MA$_{2}$Z$_{4}$ (M = transition metals; A = IVA elements; Z = VA elements)[Y.-L. Hong \emph{et al.}, Science \textbf{369}, 670 (2020)] family has led another advance for facilitating and harnessing magnetism in low-dimensional materials. However, only Cr and V based MA$_{2}$N$_{4}$ compounds exhibit intrinsic magnetism yet with unsatisfied magnetic ordering temperature. Herein, we identify a stable ferromagnetic number of this family, i.e., VGe$_{2}$Z$_{4}$ monolayer, by means of first-principles calculations. It is found that the magnetic configuration sustains under both compression and tensile uniaxial in-plane strain, and the former can act as a positive modulator to enhance magnetic ordering temperature ($T_{\rm{C}}$). Electronic structure calculations reveal a large band gap in the spin down channel while band-gapless in the spin up channel, an impressive near-half-metallic character, which is a favorable candidate for spintronic device.
\end{abstract}

\maketitle


\maketitle
Benefiting from the advantages of high speed, high integration density and high power transformers, two dimensional (2D) magnetism has been endowed with great promise for nanoscale spintronics\cite{Hong-science-2020-369-670,Novoselov-PNAS-2005-102-10451,Lee-Science-2008-321-385}. While exposed with larger surface area and dangle bonds, 2D materials exhibit extraordinary physical and chemical properties, which are intimately dependent on their local environment. In reverse, new phenomena and unusual functionality can be facially modulated in 2D materials compared with their bulk counterparts. Versatile measurements, such as strain\cite{Wu-APL-2021-118-113102,Ai-PCCP-2021-23-3144,Guo-JMCC-2021-9-2464,Cui-PRB-2021-103-085421,Zhong-PRB-2021-103-085142,Peng-PRB-2014-90-085402}, carrier doping\cite{Yang-ApplSurfSci-2020-524-146490}, deficiency\cite{Luo-JMCA-2021-9-15217}, surface modification\cite{Nair-NatPhys-2012-8-199}, and external field etc\cite{Rao-NanoLett-2012-12-1210}. Among them, strain modulation is the most effective approach in view of the structural flexibility of low dimensional materials which tends to deform. As a typical example, chromium trihalides (CrX$_{3}$, X=Cl, Br and I) with intrinsic magnetism, a compression strain as large as 6\% can drive a ferromagnetism (FM) to antiferromagnetism (AFM) transition\cite{Xu-RSCAdv-2019-9-23142}, and a 2.4\% tensile strain can improve magnetic ordering temperature for CrCl$_{3}$ and CrBr$_{3}$ monolayers\cite{Zheng-Nanoscale-2018-10-14298}.

It is known that 2D materials can be composed by different number of atomic layers (defined as n). Taking the well-known graphene as an example, the C atoms in the six-membered ring are within one single layer, i.e., n=1\cite{Lee-Science-2008-321-385,Kuzmenko-PRL-2008-100-117401}. While in the 2H-MoS$_{2}$ and the aforementioned CrX$_{3}$ (X=Cl, Br and I), the metal elements are sandwiched in the inner layer by the outer layers of non-metal atoms, wherein n is equal to 3. Growing to n=4, the $\alpha$-InSe has been extensively studied, holding great potential in optic field. Very recently, another success has been achieved to combine the n=3 and n=4 materials in one. In details, MoSi$_{2}$N$_{4}$ and WSi$_{2}$N$_{4}$  monolayers have been experimentally synthesized via the chemical vapor deposition method, creating a new family of 2D materials with n=7. Taking MoSi$_{2}$N$_{4}$ as an example, its geometric structure can be identified as an intercalated architecture of MoN$_{2}$ (n=3 with 2H-MoS$_{2}$ symmetry) and Si$_{2}$N$_{2}$ (n=4 with $\alpha$-InSe type).

Since reported, the MA$_{2}$Z$_{4}$ (M= transition metals; A=IVA elements; Z=VA elements) family, as another milestone, has sparked a wide range of intriguing following studies. Not only the family has been enriched by predictions of new members such as MA$_{2}$Z$_{4}$ (M=Ti, V, Cr; A=Si; Z=N, P, As), but also compositional/structural tailoring has been proposed to satisfy practical requirements \cite{Zhong-PRB-2021-103-085142,Bian-arXiv:2012.04162,Bafekry-NewJChem-2021-45-8291,Li-Nanoscale-2021-13-8038,Akanda-PRB-2020-102-224414,Wang-NatCommun-2021-12-2361,Chen-ChemEurJ-2021-27-9925,Guo-PCCP-2020-22-28359,Li-AdP-2021-533-2100273,
Akanda-APL-2021-119-052402,Feng-APL-2022-120-092405}. In the contribution, Zhong et al revealed that under external strain, MSi$_{2}$N$_{4}$ (M=Ti, Cr, Mo) undergoes a phase transition which is attributed to the asymmetric charge transfer\cite{Zhong-PRB-2021-103-085142}. Effect of O doping and N vacancy on structural, electronic and magnetic properties of MoSi$_{2}$N$_{4}$ has been systematically explored by Bian et al\cite{Bian-arXiv:2012.04162}. In order to improve the absorption of light spectra, a heterostructure of MoS$_{2}$/MoSi$_{2}$N$_{4}$ is proposed, whose work function is smaller than that of either MoS$_{2}$ or MoSi$_{2}$N$_{4}$\cite{Bafekry-NewJChem-2021-45-8291}. When interfaced with magnetic material, interfacial Dzyaloshinskii-Moriya Interaction (iDMI) which is beneficial to achieve spintrionic devices\cite{Akanda-PRB-2020-102-224414}. More interestingly, a Janus structure is designed for MSi$_{2}$C$_{x}$N$_{4-x}$ (M=Cr, Mo, and W; x=1 and 2) with controllable electronic and magnetic character\cite{Li-Nanoscale-2021-13-8038}.

The potential applications of MA$_{2}$Z$_{4}$ involves mechanical engineering, electronics, optic devices, electrocatalysts (N$_{2}$ reduction and hydrogen evolution), etc. However, only Cr and V based candidates can achieve ferromagnetism, yet with unsatisfied magnetic ordering temperature ($T_{\rm{C}}$). Specifically, the $T_{\rm{C}}$ of CrSi$_{2}$CN$_{3}$ is as low as 123 K\cite{Li-Nanoscale-2021-13-8038}. In one of our previous work, FM interaction can be enhanced by in-plane biaxial strain in VSi$_{2}$N$_{4}$, giving rise to a better $T_{\rm{C}}$ (285 K under 5\% tensile stain)\cite{Wang-NatCommun-2021-12-2361}, still not reaching room-temperature. To be competent in industrial grade, it is thus highly desirable to identify new component and effective approach to reach robust magnetic ordering temperature. Herein, VGe$_{2}$N$_{4}$, a sister compound of VSi$_{2}$N$_{4}$, has been systematically studied under strain. It is found that FM phase is stable under tensile/compressive strain and $T_{\rm{C}}$ can be improved with the latter modulation, which reaches 325 K from Monte Carlo simulation.

\begin{figure*}[htp!]
\centerline{\includegraphics[width=0.9\textwidth]{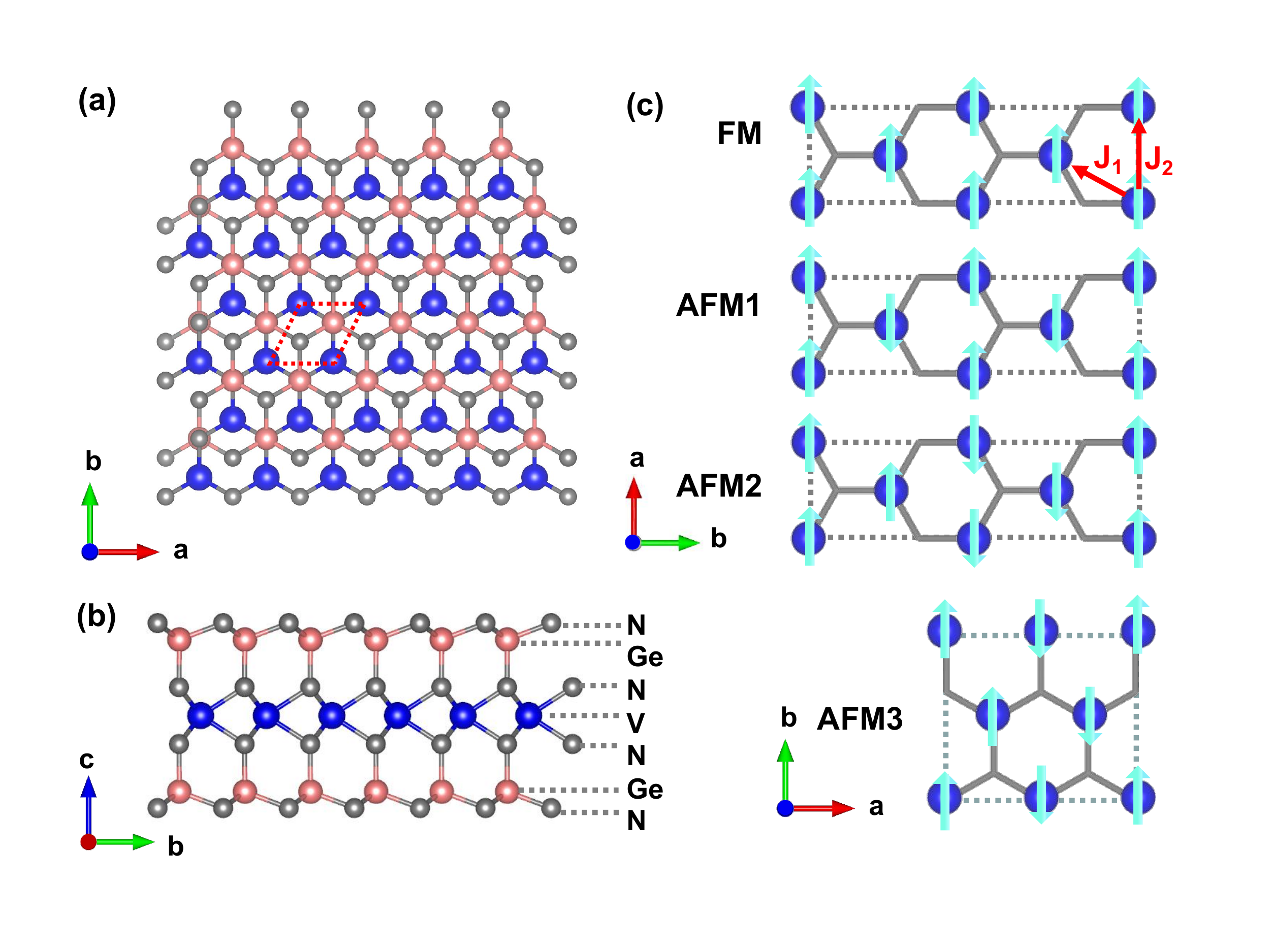}}
\caption{{\bf Crystal structure of layered VGe$_{2}$N$_{4}$ (a) top view and (b) side view of the septuple layer. (c) Magnetic configurations designed for the VN$_{2}$ layer. The directions of the arrow indicate the spin orientations.}
\label{fig:latt}}
\end{figure*}

First-principles calculations were performed as implemented in Vienna ab initio simulation package (VASP)\cite{Kresse-PRB-1996-54-11169,Kresse-CMS-1996-6-15,Kresse-PRB-1993-47-558,Kresse-PRB-1994-49-14251}. Structural relaxations and electronic properties were calculated with the projector augmented-wave (PAW) method\cite{Kresse-PRB-1999-59-1758,Bloch-PRB-1994-50-17953}. The exchange correlation energy was treated by the generalized gradient approximation (GGA - PBE)\cite{Perdew-PRL-1996-77-3865}. GGA+U method\cite{Liechtenstein-PRB-1995-52-R5467} was applied to describe the correlated V 3d electrons and U$_{eff}$ was set as 3 eV, in accordance with previous study\cite{Cui-PRB-2021-103-085421}. The plane wave cut-off energy was set to be 500 eV. The k-mesh was sampled by a $9 \times 9 \times 1$ mesh for structural optimization and a finer $13 \times 13\times 1$ was applied for electronic structure calculations. The vdW interaction correlation was considered by using density functional theory force-field approach (DFT-D2)\cite{Grimme-JCP-2010-132-154104,Grimme-JCC-2006-27-1787}. The energy convergence criterion was $1.0 \times 10^{-8} $eV, and the force was set for $1.0 \times 10^{-2} $ eV/\AA. The vacuum of 15 \AA \/ is set to avoid the interlayer interaction. Phonon dispersion was calculated using Phonopy package and a $3 \times 3$ supercell was built for force calculations\cite{Togo-2015-108-1}. Magnetic ordering temperature was estimated by Monte Carlo simulation within Heisenberg model\cite{Liu-ApplSurfSci-2019-480-300}.

\begin{figure*}[htp!]
\centerline{\includegraphics[width=1.0\textwidth]{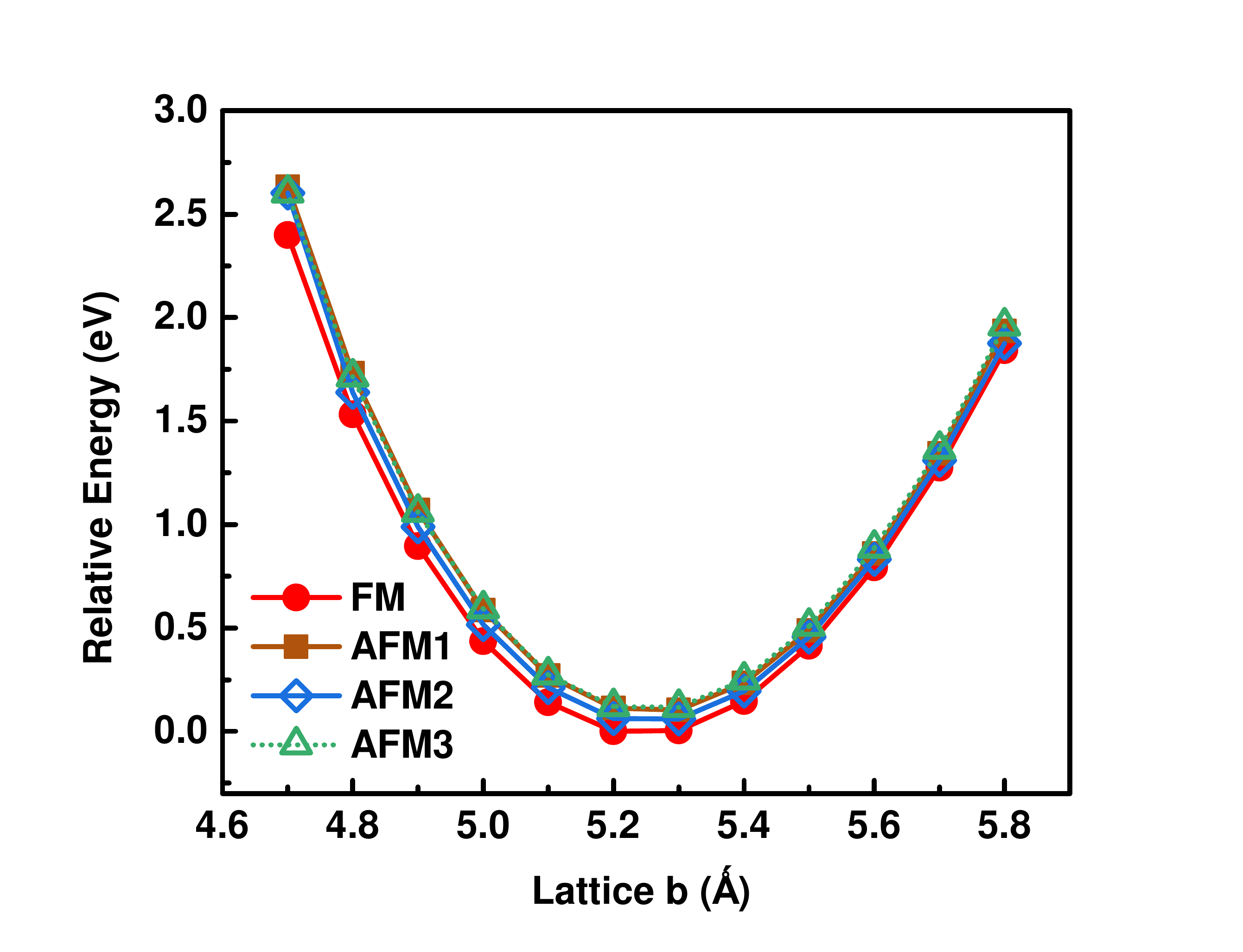}}
\caption{{\bf  Uniaxial strain along b direction for VGe$_{2}$N$_{4}$.}
\label{fig:es}}
\end{figure*}

The VGe$_{2}$N$_{4}$ features with a triangular lattice (space group P-6m2, No. 187) and consists of septuple atomic layers with the sequence of N-Ge-N-V-N-Ge-N, which can be identified as a VN$_{2}$ layer sandwiched by two Ge-N bilayers. (Fig. 1a-b) The optimized in-plane lattice $a$ is 3.008 \AA, larger than that of VSi$_{2}$N$_{4}$ (2.880 \AA)\cite{Cui-PRB-2021-103-085421}, which is understandable in view of the increased radii of Ge. This trend is further verified from our calculations for other four V based counterparts (VA$_{2}$Z$_{4}$ with A= Ge and Si; Z = P and As, Fig.S1). It is seen that the heavier the element for Z, the larger the lattice constant is, and the influence is more obvious for out-of-plane lattice $c$. In the VN$_{2}$ layer, each V atom is coordinated with six N atoms, forming a perfect triangular prism. The V-N bond length is 2.070 \AA, slightly larger than that in VSi$_{2}$N$_{4}$ (~2.040 \AA)\cite{Li-AdP-2021-533-2100273}, which produces different magnetic evolution under strain. In the upper and bottom outer Ge-N layer, a Ge atom coordinates to four N atoms, which leads to tetrahedron environment. The structural stability is verified through the calculated phonon dispersion (Fig.S2).

\begin{figure*}[htp!]
\centerline{\includegraphics[width=0.9\textwidth]{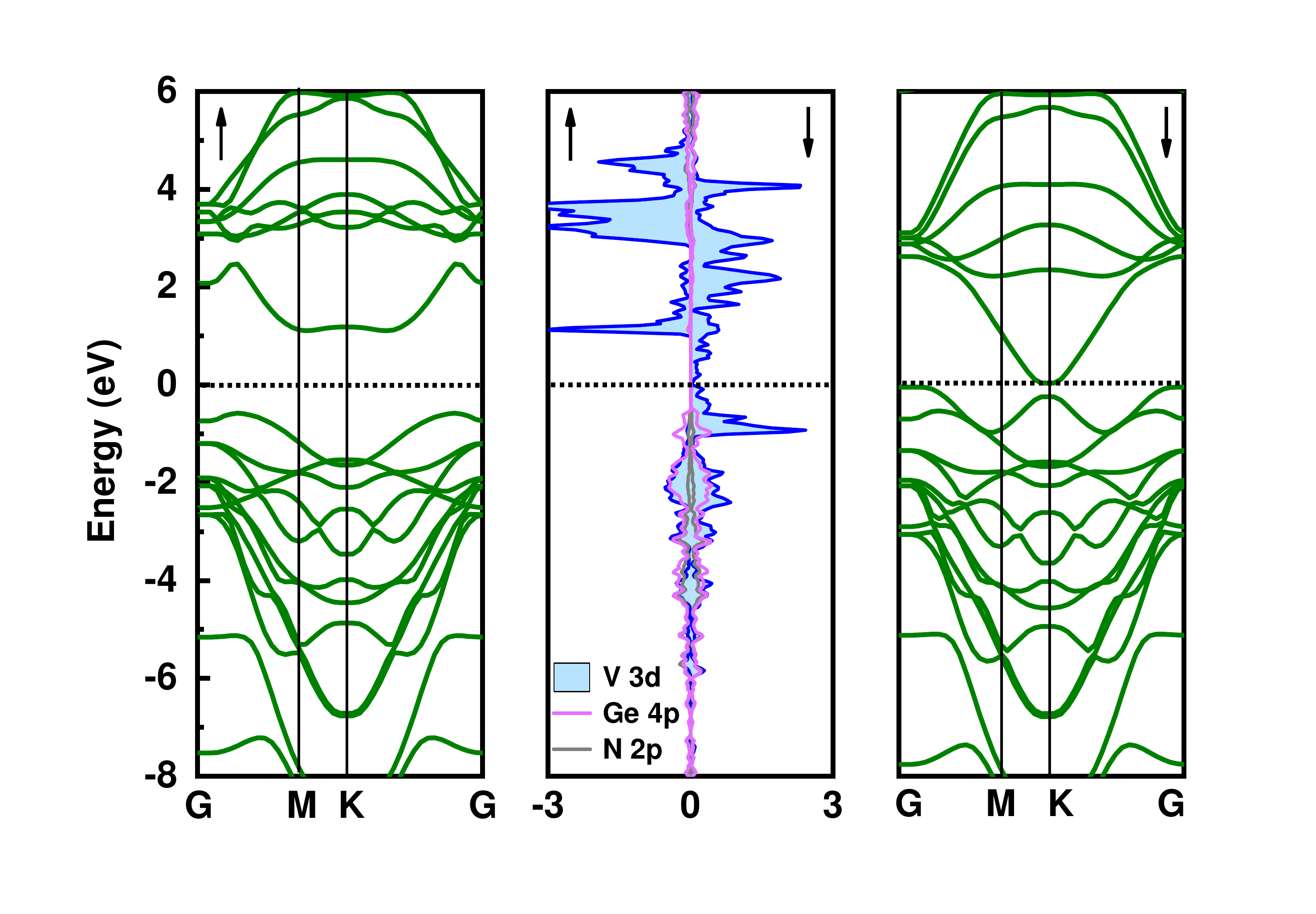}}
\caption{ {\bf Band structure and partial density of states for VGe$_{2}$N$_{4}$.}
\label{fig:elf}}
\end{figure*}

In the equilibrium lattice (V$_{0}$), one transition metal is shared by three rings and in each ring, there are two nearest neighbors, giving rise to six identical V-V interactions. However, when uniaxial strain is loaded, the resultant V-N bonds produce distinct V-V interactions (J$_{1}$ and J$_{2}$, respectively), as shown in Fig.1c. Accordingly, four different magnetic configurations (FM, AFM1-AFM3) are designed in a $\sqrt{2}\times \sqrt{2}$ supercell (Fig.1c). In-plane uniaxial strain along $b$ direction is considered. In the equilibrium lattice, the spin magnetic moment of V parallels with each other, exhibiting FM character. The spin magnetic moment is 0.988 $\mu_{B}$ in GGA method, which is enhanced to be 1.088 $\mu_{B}$ when U is involved. This value is much lower than the theoretical value for V$^{3+}$ (3$d^{2}$) and underlines an itinerant character even in GGA+U approach. For simplicity, we mainly discuss the results from GGA+U in the following sections. It is seen that FM configuration sustains stable under both tensile (16\% at most, corresponding lattice $b$ = 5.8 \AA) and compression strain (up to 10\% at $b$ = 4.7 \AA) considered (Fig.2).

\begin{figure*}[htp!]
\centerline{\includegraphics[width=0.9\textwidth]{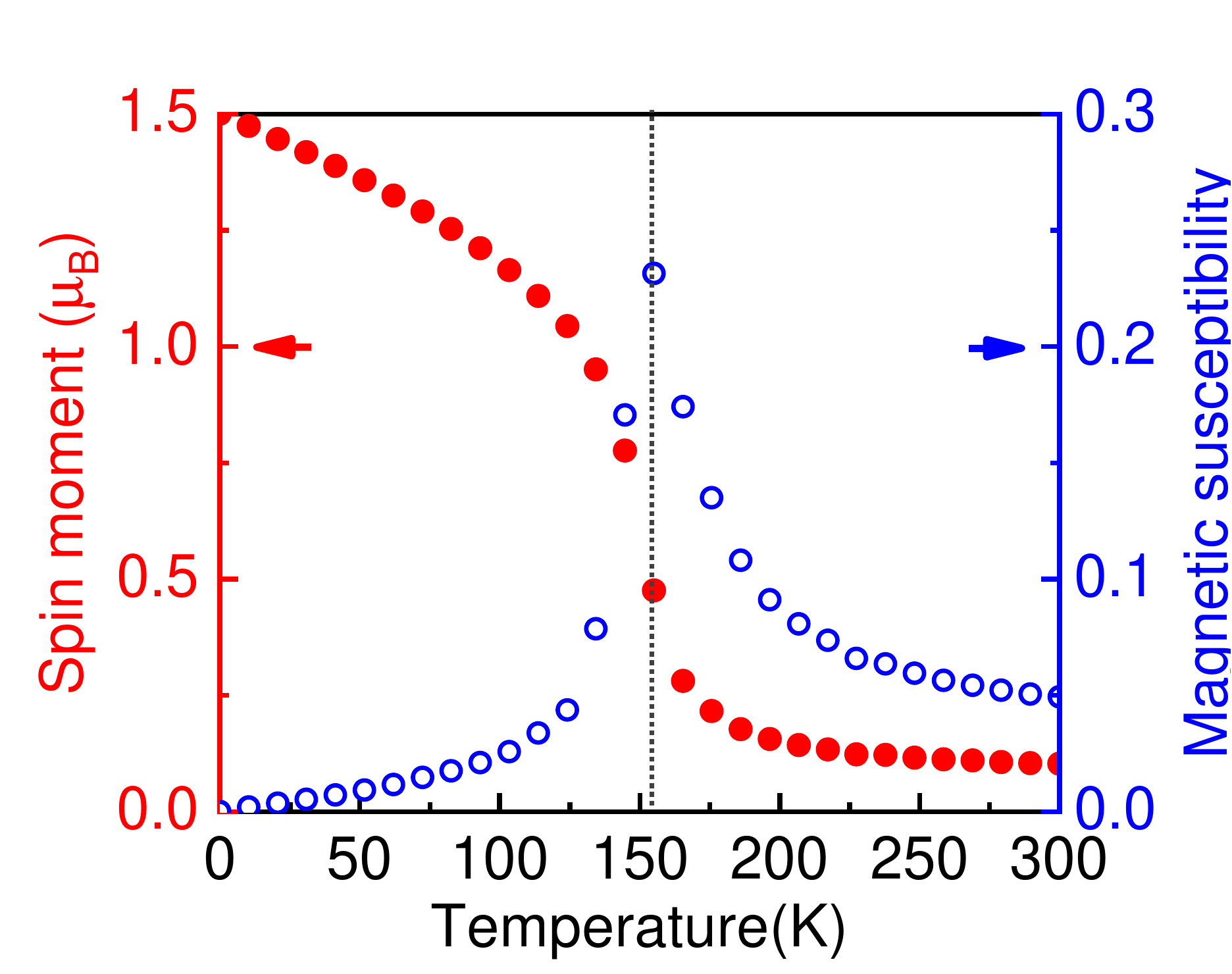}}
\caption{{\bf Average magnetic moment and susceptibility from Monte Carlo simulations for VGe$_{2}$N$_{4}$.}
\label{fig:cd}}
\end{figure*}

The calculated electronic structure shows asymmetry with different behaviors in spin up and spin down channels (Fig.3). In equilibrium lattice, a band gap as large as 1.69 eV is formed within V 3$d$ and Ge 4$p$ orbitals in the spin down channel, while in the opposite spin channel, both the valence band maximum and conduction band minimum touch the Fermi level (E$_{F}$), forming a unique spin gapless state, which survives under tensile strain and small compression uniaxial strain (Fig S3-S12). However, large compression uniaxial strain produces a small band gap in the spin down direction (Fig. S13-S14). Taking the case of 10\% compression strain as an example, the band gap in the spin down channel is 0.54 eV.

\begin{figure*}[htp!]
\centerline{\includegraphics[width=0.9\textwidth]{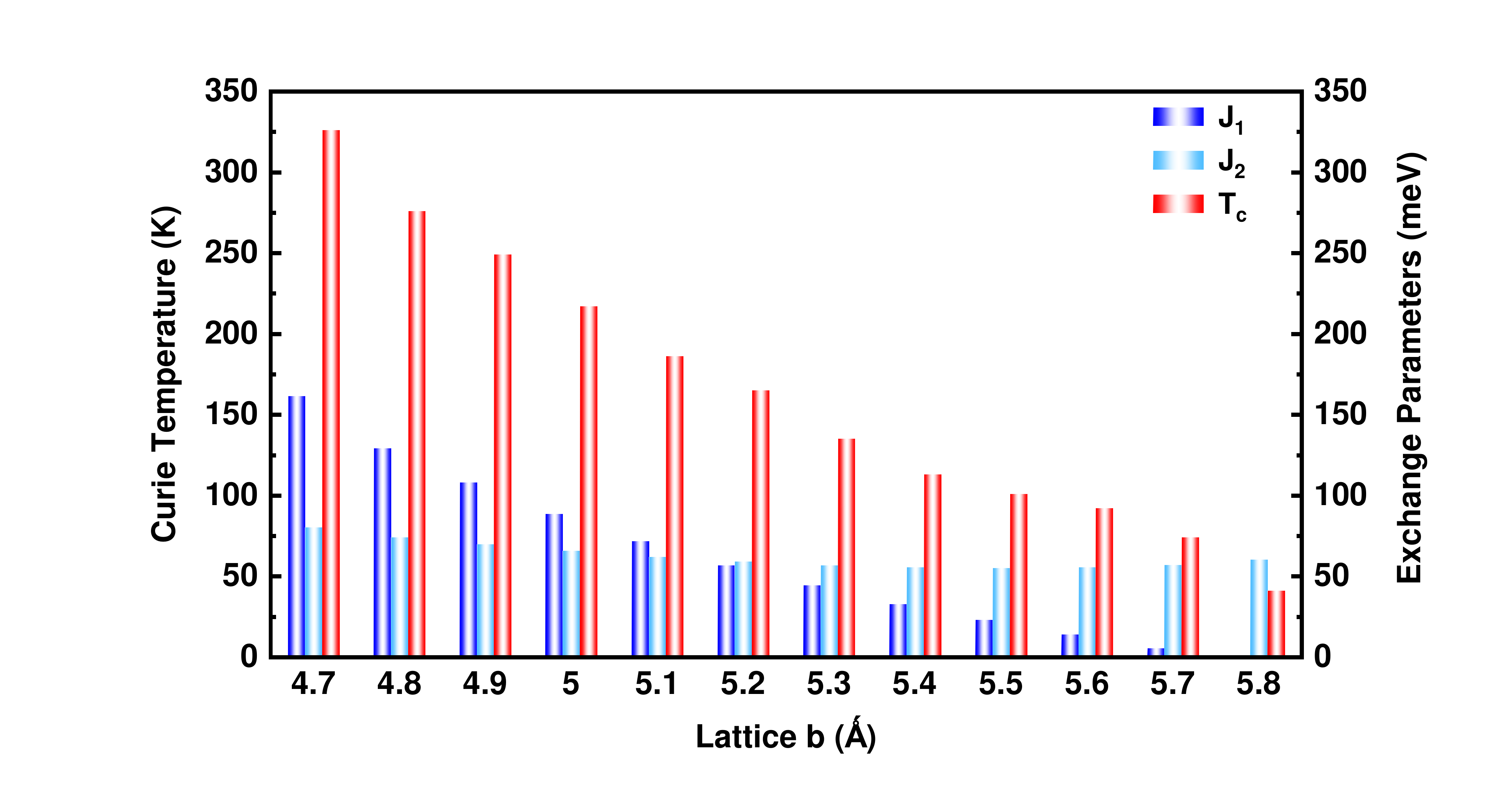}}
\caption{{\bf Calculated magnetic exchange parameters and Curie temperatures ($T_{\rm{C}}$) for VGe$_{2}$N$_{4}$ under uniaxial strain.}
\label{fig:cd}}
\end{figure*}

Magnetic exchange parameters have been calculated to evaluate the magnetic ordering temperature. In equilibrium lattice (V$_{0}$), each transition metal has six nearest neighbors, giving rise to six identical magnetic interaction(J), which can be derived from comparing the energy from FM to AFM1 (6J = E$_{AFM}$ - E$_{FM}$). When uniaxial strain is considered, distinction in lattice generates different magnetic interaction, which are labeled as J$_{1}$ and J$_{2}$ (Fig.1). Using the calculated magnetic parameters, $T_{\rm{C}}$ has been estimated from Monte Carlo simulation. It is found that $T_{\rm{C}}$ is up to 154 K for the VGe$_{2}$N$_{4}$ in ambient conditions(Fig.4). Deviating from the equilibrium, the magnetic exchange parameter J$_{1}$ is enhanced in compression strain and decreased in tense strain, while the change of J$_{2}$ is small (Fig.5). The enhancement of J$_{1}$ can be understood from the structure variation wherein the compression strain shorten the V-N bond length and gave a V-N-V bond angle approaching to 90$^{0}$, contributing to improve the 90$^{0}$ FM interaction. Meanwhile, there are four paths for J$_{1}$ yet only two for J$_{2}$. As a result, the variation of $T_{\rm{C}}$ with strain is the alike that of J$_{1}$. It is favorable that when the compression strain is as large as 16\%, $T_{\rm{C}}$ reaches 325 K(Fig.5), over room temperature. Moreover, magnetic anisotropy energy (MAE) has been calculated considering the easy magnetization along [100] (in-plane) and [001] (out-of-plane). Small MAE is obtained for VGe$_{2}$N$_{4}$ with/without strain (Table S1), which gives rise to negligible anisotropic exchange parameter and ignorable influence on $T_{\rm{C}}$ (Fig. S16).

In summary, the electronic and magnetic properties of the septuple atomic layer VGe$_{2}$N$_{4}$ have been studied through first-principles calculations. VGe$_{2}$N$_{4}$ exhibits ferromagnetic nature which retains under compression and tensile uniaxial strain. It is found that magnetic ordering temperature can be improved with the help of compression strain which strengthens the nearest magnetic exchange interaction. Our study not only predicts a good candidate in this family but also initiates a starting point for further pursuit of more 2D materials with high-performance of magnetism.

\section{Supplementary Material}
See the supplementary material for one table and sixteen figures: One table shows magnetic anisotropy energy for VGe$_{2}$N$_{4}$ under strain: Fig.S1 shows lattice parameters for VGe$_{2}$N$_{4}$, VGe$_{2}$P$_{4}$, VGe$_{2}$As$_{4}$, VSi$_{2}$P$_{4}$ and VSi$_{2}$P$_{4}$; Fig.S2 shows  calculated phonon dispersion curves for VGe$_{2}$N$_{4}$; Fig. S3-S14 show Band structure and partial density of states for VGe$_{2}$N$_{4}$ with different b(5.8\AA,5.7\AA,5.6\AA,5.5\AA,5.4\AA,5.3\AA,5.2\AA,5.1\AA,5.0\AA,4.9\AA,4.8\AA,4.7\AA).
FIG.S15 shows "average magnetic moment and susceptibility from Monte Carlo simulations
for VGe$_{2}$N$_{4}$ under strain". FIG.S16 shows "average magnetic moment and susceptibility from Monte Carlo simulations
for VGe$_{2}$N$_{4}$ with anisotropic Heisenberg model".


This work was supported by the Natural Science Foundation of Hebei Province (No. A2019203507). The authors thank the High Performance Computing Center of Yanshan University.

\section{DATA AVAILABILITY}
The data that support the findings of this study are available from the corresponding author upon reasonable request.


\bibliography{apssamp}

\section{References}
\begin{thebibliography}{99}

\bibitem{Hong-science-2020-369-670}
Y.-L. Hong, Z. Liu, L. Wang, T. Zhou, W. Ma, C. Xu, S. Feng, L. Chen, M.-L. Chen, D.-M. Sun, X.-Q. Chen, H.-M. Cheng, W. Ren, Science, \textbf{369}, 670 (2020).

\bibitem{Novoselov-PNAS-2005-102-10451}
K.~S. Novoselov, D. Jiang, F. Schedin, T.-J. Booth, V.~V. Khotkevich, S.~V. Morozov, A.~K. Geim, Proc. Natl. Acad. Sci. \textbf{102}, 10451 (2005).

\bibitem{Lee-Science-2008-321-385}
C. Lee, X. Wei, J.~W. Kysar, J. Hone, Science \textbf{321}, 385 (2008).

\bibitem{Wu-APL-2021-118-113102}
Q.-Y. Wu, L.-M. Cao, Y.-S. Ang, L.-K. Ang, Appl. Phys. Lett.\textbf{118}, 113102 (2021).

\bibitem{Ai-PCCP-2021-23-3144}
 H.~Q. Ai, D. Liu, J.~Z. Geng, S.~P. Wang, K.~H. Lo, H. Pan, Phys. Chem. Chem. Phys. \textbf{23}, 3144 (2021).

\bibitem{Guo-JMCC-2021-9-2464}
 S.~D. Guo, W.~Q. Mu, Y. -T. Zhu, R.~Y. Hana, W.~C Ren, J. Mater. Chem. C \textbf{9}, 2464 (2021).

\bibitem{Cui-PRB-2021-103-085421}
Q.~R. Cui, Y. M. Zhu, J. H. Liang, P. Cui, H.~X. Yang, Phys. Rev. B \textbf{103}, 085421 (2021).

\bibitem{Zhong-PRB-2021-103-085142}
H.~X. Zhong, W.~Q. Xiong, P.~F. Lv, J. Yu, S.~J. Yuan, Phys. Rev. B \textbf{103}, 085142 (2021).

\bibitem{Peng-PRB-2014-90-085402}
X. Peng, Q.Wei, A. Copple, Phys. Rev. B \textbf{90}, 085402 (2014).

\bibitem{Yang-ApplSurfSci-2020-524-146490}
J. Yang, J. Wang, Q. Liu, R. Xu, Y. Li, M. Xia, Z. Li, F. Gao, Appl. Surf. Sci. \textbf{524}, 146490 (2020).

\bibitem{Luo-JMCA-2021-9-15217}
Y. Luo, M.~Y. Li, Y.~X. Dai, X.~L. Zhang, R.~Q. Zhao, F. Jiang, C.~Y. Ling, Y.~C. Huang, J. Mater. Chem. A \textbf{9}, 15217 (2021).

\bibitem{Nair-NatPhys-2012-8-199}
R. Nair, M. Sepioni, I.~L. Tsai, O. Lehtinen, J. Keinonen, A. Krasheninnikov, T. Thomson, A. Geim, I. Grigorieva, Nat. Phys. \textbf{8}, 199 (2012).

\bibitem{Rao-NanoLett-2012-12-1210}
S. Rao, S. Jammalamadaka, A. Stesmans, V. Moshchalkov, J. van Tol, D. Kosynkin, A. Higginbotharn Duque, J. Tour, Nano Lett. \textbf{12}, 1210 (2012).

\bibitem{Xu-RSCAdv-2019-9-23142}
C. Xu, J. Zhang, M. Guo, L. Wang, RSC Adv. \textbf{9}, 23142 (2019).

\bibitem{Zheng-Nanoscale-2018-10-14298}
 F. Zheng, J. Zhao, Z. Liu, M. Li, M. Zhou, S. Zhang, P. Zhang, Nanoscale \textbf{10}, 14298 (2018).

\bibitem{Kuzmenko-PRL-2008-100-117401}
 A.~B. Kuzmenko, E. van Heumen, F. Carbone, D. van der Marel, Phys. Rev. Lett. \textbf{100}, 117401 (2008).

\bibitem{Bian-arXiv:2012.04162}
 Y.~T. Bian, G.~H. Liu, S.~H Qian, X.~X. Ding, H.~X. Liu, arXiv:2012.04162.

 \bibitem{Bafekry-NewJChem-2021-45-8291}
 A. Bafekry, M. Faraji, A. A. Ziabari, M. M. Fadlallah, C. V. Nguyen, M. Ghergherehchi, S. A. H. Feghhi, New J. Chem. \textbf{45}, 8291 (2021).

\bibitem{Akanda-PRB-2020-102-224414}
M. R. K. Akanda, I. J. Park, R. K. Lake, Phys. Rev. B \textbf{102}, 224414 (2020).

\bibitem{Li-Nanoscale-2021-13-8038}
 B.~W. Li, J.~Z. Geng, H.~Q. Ai, Y.~C. Kong, H.~Y. Bai, K.~H. Lo, K.~W. Ng, Y. Kawazoec, H. Pan, Nanoscale \textbf{13}, 8038 (2021).

\bibitem{Wang-NatCommun-2021-12-2361}
L. Wang, Y.~P. Shi, M.~F. Liu, A. Zhang, Y.~L. Hong, R.~H. Li, Q. Gao, M.~X. Chen, W.~C. Ren, H.~M. Cheng, Y.~Y. Li, X.~Q. Chen, Nat. Commun. \textbf{12}, 2361 (2021).

\bibitem{Chen-ChemEurJ-2021-27-9925}
J. Chen, Q. Tang, Chem. Eur. J. \textbf{27}, 9925 (2021).

\bibitem{Guo-PCCP-2020-22-28359}
 S.~D. Guo, W.~Q. Mu, Y.~T. Zhu, X.~Q. Chen, Phys. Chem. Chem. Phys. \textbf{22}, 28359 (2020).

\bibitem{Li-AdP-2021-533-2100273}
 Y.Li, J. Wang, G.~C. Yang, Y. Liu, Annalen Der Physik \textbf{533}, 2100273 (2021).

\bibitem{Akanda-APL-2021-119-052402}
M. R. K. Akanda, R. K. Lake, Appl. Phys. Lett. \textbf{119}, 052402 (2021).

\bibitem{Feng-APL-2022-120-092405}
Y. L. Feng, Z. L. Wang, X. Zuo, G. Y. Gao Appl. Phys. Lett. \textbf{120}, 092405 (2022).

\bibitem{Kresse-PRB-1996-54-11169}
G. Kresse, J. Furthmller, Phys. Rev. B \textbf{54}, 11169 (1996).

\bibitem{Kresse-CMS-1996-6-15}
G. Kresse, J. Furthmller, Comput. Mater. Sci. \textbf{6}, 15 (1996).

\bibitem{Kresse-PRB-1993-47-558}
G. Kresse, J. Hafner, Phys. Rev. B \textbf{47}, 558 (1993).

\bibitem{Kresse-PRB-1994-49-14251}
 G. Kresse, J. Hafner, Phys. Rev. B \textbf{49}, 14251 (1994).

\bibitem{Kresse-PRB-1999-59-1758}
 G. Kresse, D. Joubert, Phys. Rev. B \textbf{59}, 1758 (1999).

\bibitem{Bloch-PRB-1994-50-17953}
 P.~E. Blochl, Phys. Rev. B \textbf{50}, 17953 (1994).

\bibitem{Perdew-PRL-1996-77-3865}
J.~P. Perdew, K. Burke, M. Ernzerhof, Phys. Rev. Lett. \textbf{77}, 3865 (1996).

\bibitem{Liechtenstein-PRB-1995-52-R5467}
A.~I. Liechtenstein, V. I. Anisimov, J. Zaanen, Phys. Rev. B \textbf{52}, R5467 (1995).

\bibitem{Grimme-JCP-2010-132-154104}
S. Grimme, J. Antony, S. Ehrlich, H. Krieg, J. Chem. Phys. \textbf{132}, 154104 (2010).

\bibitem{Grimme-JCC-2006-27-1787}
S. Grimme, J. Comput. Chem. \textbf{27}, 1787 (2006).

\bibitem{Togo-2015-108-1}
 A. Togo, I. Tanaka, Scr. Mater. \textbf{108}, 1 (2015).

\bibitem{Liu-ApplSurfSci-2019-480-300}
L. Liu, X. Ren, J.~H. Xie, B. Cheng, W.~K. Liu, T.~Y. An, H.~W. Qin, J.~F. Hu, Appl. Surf. Sci. \textbf{480}, 300 (2019).

\end{thebibliography}

\end{document}